\documentclass{aa}  

\usepackage{graphicx}
\usepackage{txfonts}

\begin{document}

   \title{The eROSITA Final Equatorial-Depth Survey (eFEDS):}

   \subtitle{The first archetypal Quasar in the feedback phase discovered by eROSITA }

   \author{
M. Brusa \inst{1,2} 
\and T. Urrutia \inst{3}
\and Y. Toba \inst{4,5,6}
\and J. Buchner   \inst{7}
\and J.-Y. Li \inst{8,9,10}
\and T. Liu  \inst{7}
\and M. Perna \inst{11,12}
\and M. Salvato \inst{7,13}
\and A. Merloni  \inst{7}
\and B. Musiimenta \inst{1,2}
\and K. Nandra   \inst{7}
\and J. Wolf \inst{7,13}
\and R. Arcodia  \inst{7}
\and T. Dwelly   \inst{7}
\and A. Georgakakis   \inst{14}
\and A. Goulding   \inst{15}
\and Y. Matsuoka   \inst{6}
\and T. Nagao \inst{6}
\and M. Schramm \inst{16}
\and J.D. Silverman \inst{9}
\and Y. Terashima \inst{17}
      }

   \institute{Dipartimento di Fisica e Astronomia "Augusto Righi", Universit\`a di Bologna,  via Gobetti 93/2,  40129 Bologna, Italy 
\and INAF - Osservatorio di Astrofisica e Scienza dello Spazio di Bologna, via Gobetti 93/3,  40129 Bologna, Italy 
\and Leibniz-Institut f\"ur Astrophysik Potsdam (AIP). An der Sternwarte 16. 14482 Potsdam, Germany 
\and Department of Astronomy, Kyoto University, Kitashirakawa-Oiwake-cho, Sakyo-ku, Kyoto 606-8502, Japan
\and Academia Sinica Institute of Astronomy and Astrophysics, 11F of Astronomy-Mathematics Building, AS/NTU, No.1, Section 4, Roosevelt Road, Taipei 10617, Taiwan 
\and Research Center for Space and Cosmic Evolution, Ehime University, 2-5 Bunkyo-cho, Matsuyama, Ehime 790-8577, Japan 
\and Max Planck Institut f\"ur Extraterrestrische Physik, Giessenbachstrasse 1, 85748 Garching bei M\"unchen, Germany 
\and CAS Key Laboratory for Research in Galaxies and Cosmology, Department of Astronomy, University of Science and Technology of China, Hefei 230026, China
\and Kavli Institute for the Physics and Mathematics of the Universe (WPI), The University of Tokyo, Kashiwa, Chiba 277-8583, Japan
\and School of Astronomy and Space Science, University of Science and Technology of China, Hefei 230026, China
\and Centro de Astrobiología, (CAB, CSIC–INTA), Departamento de Astrof\'isica, Cra. de Ajalvir Km. 4, 28850 – Torrej\'on de Ardoz, Madrid, Spain
\and  INAF - Osservatorio Astrofisico di Arcetri, Largo Enrico Fermi 5, I-50125 Firenze, Italy
\and Exzellenzcluster ORIGINS, Boltzmannstr. 2, D-85748 Garching, Germany 
\and Institute for Astronomy and Astrophysics, National Observatory of Athens, V. Paulou and I. Metaxa, 11532, Greece
\and Department of Astrophysical Sciences, Princeton University, Princeton, NJ, 08544, USA
\and National Astronomical Observatory of Japan, Mitaka, Tokyo 181-8588, Japan
\and Graduate School of Science and Engineering, Ehime University, 2-5 Bunkyo-cho, Matsuyama, Ehime 790-8577, Japan
             }
             
   \date{Received April 15 ; accepted ...}

\titlerunning{A Quasar in the feedback phase discovered by eROSITA in eFEDS}

  \abstract{Theoretical models of galaxy-AGN co-evolution ascribe an important role for the feedback process to a short, luminous, obscured, and dust-enshrouded phase during which the accretion rate of the SMBH is expected to be at its maximum and the associated AGN-driven winds are also predicted to be maximally developed. To test this scenario, we have isolated a text-book candidate from the  {\it eROSITA} Final Equatorial-Depth Survey (eFEDS) obtained within the Performance and Verification program of the {\it eROSITA} telescope on board Spectrum R\"ontgen Gamma (SRG). From an initial catalog of 246 {\it hard} X-ray selected sources matched with the photometric and spectroscopic information available within the {\it eROSITA} and Hyper Suprime-Cam consortia,  three  candidates Quasars (QSOs) in the feedback phase have been isolated applying the diagnostic proposed in Brusa et al. (2015). Only one source (eFEDSU J091157.5+014327) has a spectrum already available (from SDSS-DR16, z=0.603) and it unambiguously shows the presence of a broad component (FWHM$\sim1650$ km/s) in the [OIII]5007 line. The associated observed L$_{[OIII]}$ is $\sim$ 2.6$\times10^{42}$ erg/s, one to two orders of magnitude larger than that observed in local Seyferts and comparable to those observed in a sample of z$\sim0.5$ Type 1 Quasars. From the multiwavelength data available we derive an Eddington Ratio (L$_{\rm bol}$/L$_{\rm Edd}$) of $\sim$0.25, and a bolometric correction in the hard X-ray band of k$_{\rm bol}\sim10$, lower than those observed for objects at similar bolometric luminosity. These properties, along with the presence of an outflow, the high X-ray luminosity and moderate X-ray obscuration (L$_{\rm X}$ $\sim10^{44.8}$ erg/s, N$_H\sim2.7\times10^{22}$ cm$^{-2}$) and the red optical color, all match the prediction of quasars in the feedback phase from merger driven models. 
  Forecasting to the full eROSITA all-sky survey with its spectroscopic follow-up, we predict that by the end of 2024 we will have a sample of few hundreds such objects at z=0.5-2.}

   \maketitle
%

\section{Introduction}

Since the  discovery of Supermassive Black Holes (SMBH, 10$^{6}$- 10$^{10}$ M$_\odot$) in the nuclei of virtually all galaxies,  and of the relations observed in the local Universe between host and BH properties, (e.g. Kormendy \& Ho 2013),  it has become clear that the formation and evolution of galaxies and the properties of these massive dark objects sitting in their centre  are profoundly coupled to each other. 
Some mechanism must have therefore linked the innermost regions, where the SMBH gravitational field dominates, 
to the larger scales, where its influence  is expected to be negligible. 
It has been proposed that gas flows in the form of energetic jets or winds play a pivotal role in this process (see King \& Pounds 2005 for a review). Their presence may regulate both accretion and ejection of material onto and from compact objects, and the gas accelerated by the radiation pressure from the accretion disc in rapidly accreting sources interacts with the host galaxy Interstellar Medium (ISM), propagating momentum and energy over wide spatial scales, providing an efficient feedback mechanism (e.g. Zubovas \& King 2012). 
The details of this coupling are key ingredients in all models of Active Galactic Nuclei (AGN) and galaxy co-evolution. 

The outflows developing in AGN host galaxies have a multiphase nature, as expected from simulations and revealed by observations  (see Cicone et al. 2018 for a recent summary).
 Although ionised outflows can be signalled by broad/shifted or asymmetric wings in the [OIII]5007 emission line as seen in integrated spectra (e.g. Mullaney et al. 2013, Brusa et al. 2015, Zakamska et al. 2016),  in the past decade high-resolution and high signal to noise Integral Field Units IFU observations (MUSE, KMOS and SINFONI, among others) of AGN winds mostly at low-redshift (e.g. Harrison et al. 2014, Venturi et al. 2018, Ramos-Almeida et al. 2019) have played a crucial role in uncovering their extent (radius) and kinematics properties (velocity, geometry), key parameters to derive the mass outflow rates ($\dot{M}_{ion}$) and energetics ($\dot{E}_{kin}$). 
 These fast moving gas components observed at kpc scale or beyond show substantial velocities (v$\sim$1000-3000 km/s) and mass outflow rates as large as 100-1000 M$_\odot$ yr$^{-1}$ (see Fiore et al. 2017, Rupke et al. 2017 for reviews and compilations).  
The main limitation in this kind of studies beyond the local Universe is still the paucity of targets bright enough in the optical/NIR band such that the underlying physics and kinematics of systemic and outflowing gas can be spatially resolved and studied in great detail (see e.g. Cresci et al. 2015, Brusa et al. 2016).

Theoretical models ascribe an important role to a radiatively driven process associated with  a short, luminous, and dust-enshrouded phase  ("blow-out" phase, e.g. Hopkins et al. 2008, Fabian 2012) during which the SMBH accretion is indeed expected to be at its maximum (high L/L$_{\rm Edd}$; e.g. Urrutia et al. 2012 where the reddest quasars show the highest Eddington ratio). This phase is subsequent to a heavily obscured, possibly Compton Thick phase of rapidly black hole growth. In the past two decades, mid-infrared and submillimeter surveys have proven to be especially efficient in selecting obscured quasars not biased against Compton Thick sources and therefore with a complete sampling of AGN including those in this first heavily obscured one (e.g. Lacy et al. 2004, Stern et al. 2005, Alexander et al. 2005, Mart{\'\i}nez-Sansigre et al. 2006, Daddi et al. 2007, Dey et al. 2008, Lanzuisi et al. 2009, Donley et al. 2012, Assef et al. 2015). However, these samples may suffer from contamination by the host-galaxy light, and 
in some cases the AGN nature needed to be validated by X-ray stacking analysis (e.g. Fiore et al. 2009) or dedicated X-ray observations (e.g. Stern et al. 2014, Piconcelli et al. 2015). Given that the blow-out phase is expected to be only mildly obscured in the X-rays (see e.g. Hopkins et al. 2005, Blecha et al. 2018), large area {\it hard} X-ray surveys with the associated high-quality, multi-wavelength data  are probably the best tool to select these very rare sources, providing at the same time information on the nuclear obscuration, a proxy for the AGN accretion rate (e.g. L$_X$/M$_\star$) and the unambiguous fingerprint of ongoing AGN activity. 

The lack of a sensitive all-sky X-ray survey essential to provide statistical significant samples of such rare obscured QSOs will be overcome in the immediate future by {\it eROSITA} (extended ROentgen Survey with an Imaging Telescope Array, Predehl et al. 2021), an X--ray telescope 
aboard the Spectrum R\"ontgen Gamma satellite and in full operation since December 2019. In addition to the most sensitive 0.2-2.3 keV all sky survey (Merloni et al. 2012), {\it eROSITA} provides the first ever true imaging all-sky survey in the hard band (2.3-5 keV), thanks to an unprecedented combination of a large field of view, spectral and angular resolution. 
In this letter, we present the discovery of an obscured quasar showing a powerful ionised outflow from {\it eROSITA} Performance and Verification (PV) phase observations, an assessment of the selection of such rare objects  in the context of available samples of objects caught in the feedback phase at similar redshifts, and predictions for the final {\it eROSITA} all sky survey. 
We adopt the cosmological parameters H$_0$ = 70 km s$^{-1}$ Mpc$^{-1}$ , $\Omega_{\rm m}$ =0.3 and $\Omega_\Lambda$=0.7 (Spergel 2003) throughout the paper. When quoting magnitudes, the AB system will be used, unless otherwise stated. We adopt a Chabrier Initial Mass Function to derive stellar masses and Star Formation Rates (SFRs) for the target and comparison samples.  Errors are given at 1 $\sigma$.
At the redshift of the source, the physical scale is 1\arcsec$\sim$6.8 kpc.  

%
%
%
\section{Parent sample, target selection and properties}

The unique {\it eROSITA} survey science capabilities have been tested with the PV program eFEDS ({\it eROSITA} Final Equatorial-Depth Survey; Brunner et al. submitted), a mini-survey reaching the average depth of the all-sky survey over $\sim$1/350 of the sky in the survey footprint of the Hyper Suprime-Cam (HSC: Miyazaki et al. 2018) Subaru Strategic Program (HSC-SSP: Aihara et al. 2018ab, 2019).

We started from the catalog of 246 X-ray point-like sources selected in the 2.3-5 keV band (Brunner et al. submitted) imposing a threshold in detection likelihood detection likelihood$>10$, corresponding to less than 10\% spurious sources in the field, as assessed on the basis of extensive simulations described in Liu et al. (submitted).
Salvato et al. (submitted) provides the counterparts identification for all point-like X-rays sources in eFEDS, a reliability flag for the association, and all the photometry for the proposed counterparts over the entire range of wavelengths from Ultra Violet (GALEX) to  Mid Infrared (WISE W4). We refer to Salvato et al. (submitted) for all the relevant details on the association process and description of the photometry, and to Nandra et al. (submitted) for a full characterisation of the hard sample in terms of its X-ray and multiwavelength properties. For this work we just note that a total of 231/246 sources in the hard sample have reliable optical identifications and the remaining 15/246 may be spurious associations. Photometry in the r-band (from Legacy Survey LS8 data) and W1 (from WISE) is available for all of them.

In order to select  QSOs candidates in the feedback phase, we applied a diagnostic similar to that proposed in Brusa et al. (2015), which was tested over the 2 deg$^2$ area of the XMM-COSMOS survey (Hasinger et al. 2007, Brusa et al. 2010).  In details, we imposed a color selection based on the NIR to optical flux ratio (r-W1$>4$) and on the X-ray to optical flux ratio (logX/O$>1$, where "X" is the 2-10 keV flux\footnote{We derived the 2-10 keV flux from the 2.3-5 keV flux tabulated in the eFEDS catalog by applying a correction factor of 2.27 assuming $\Gamma$=1.7.} and "O" is the r-band flux; see Brusa et al. 2010, 2015 for a  detailed discussion), and we isolated 3 such sources. Figure 1 shows the diagnostic diagram with all the sources from the eFEDS hard sample, and the selection locus. Solid and dashed symbols denote reliable and less reliable associations, respectively. 

   \begin{figure}
   \centering
   \includegraphics[width=9cm]{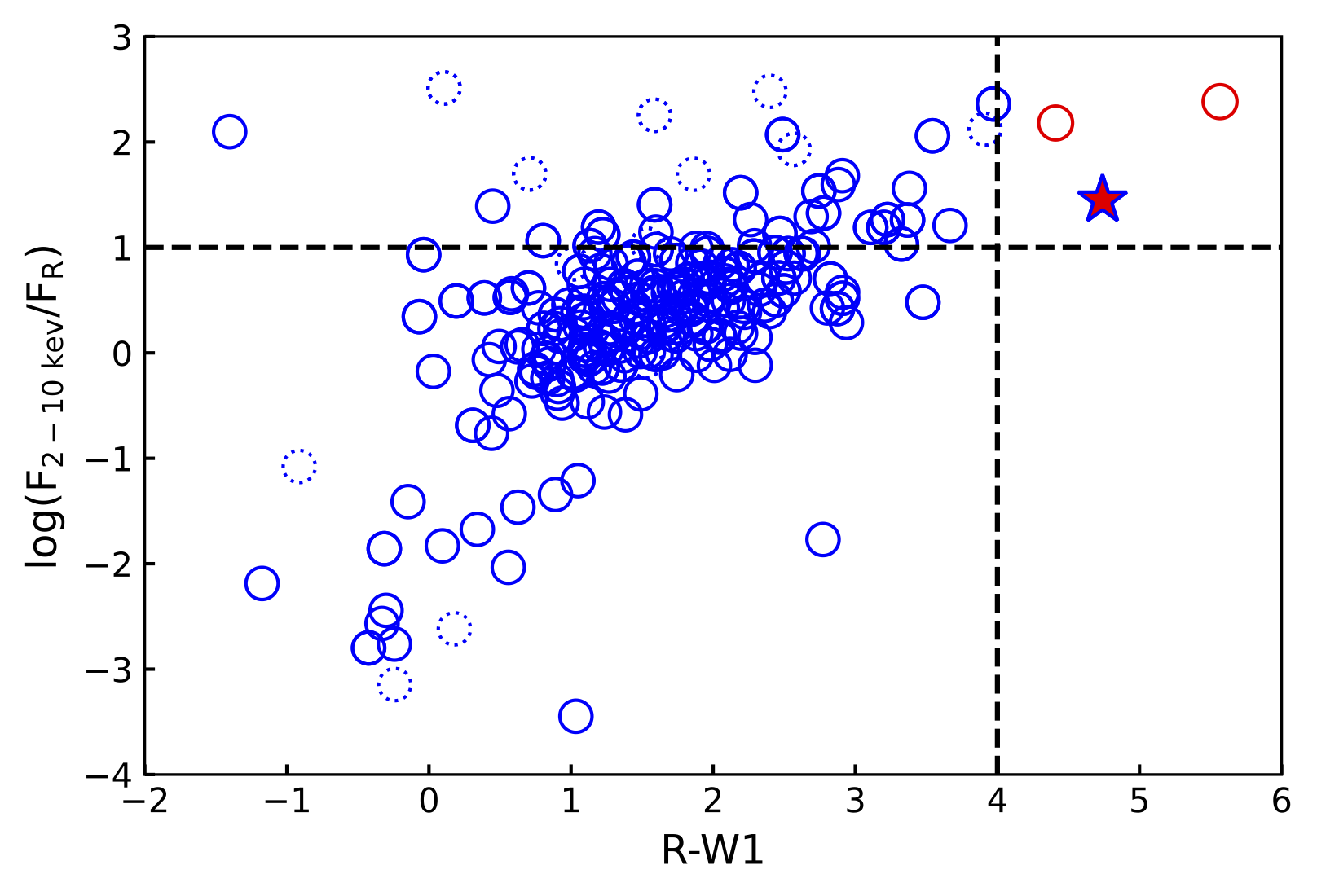}
   \caption{Diagnostic diagram used to isolate QSOs candidates in the feedback phase: X--ray to optical flux ratio vs. r-W1 colors for all the 246 eFEDS sources detected in the hard band in the eFEDS sample. Red points mark sources in the  selection locus for the "windy" QSOs, at r-W1$>4$ and logX/O$>1$ (upper right corner). Blue points mark all other sources. Sources with less reliable counterparts in the hard X-ray sample are marked with dotted lines.  
   Source XID439 is marked by a large, red star.}
              \label{FigSel}
    \end{figure}
   \begin{figure}
   \centering
    \includegraphics[width=8cm]{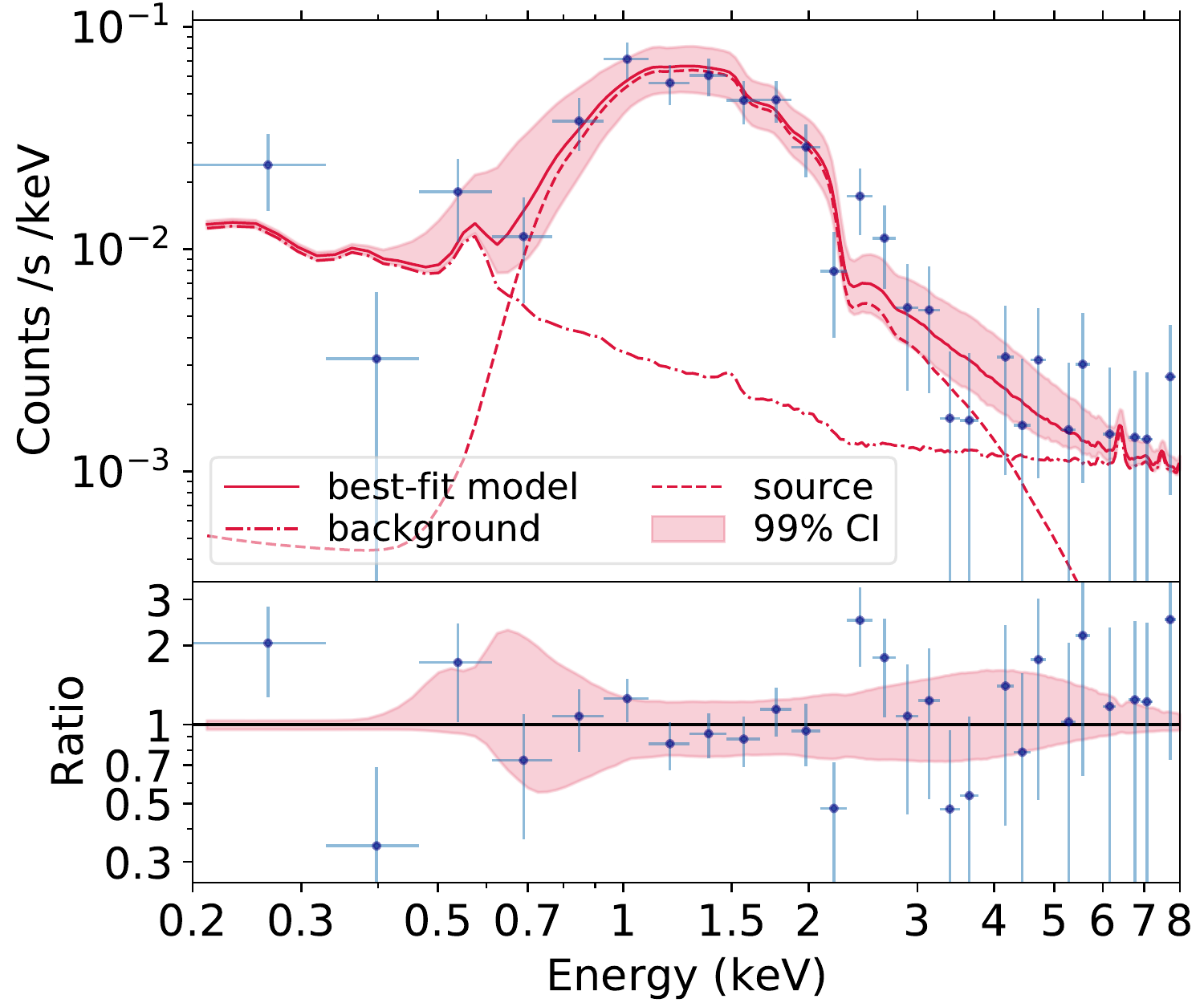} 
      \caption{{\it eROSITA} spectrum of XID439 (blue points)  with superimposed the source, background and source+background models, as labeled. {\it bottom panel:} Ratio of data with respect to the best fit model.}
         \label{FigSpe}
   \end{figure}

Of these three {\it eROSITA} objects, only one source has a spectroscopic redshift already available from SDSS DR16 (Lyke et al. 2020), eFEDSU J091157.5+014327 at z=0.603 (XID439 in the eFEDS catalog, shown as a red star in Figure~\ref{FigSel}; Nandra et al. submitted).  

\subsection{X-ray spectrum}
Bayesian spectral analysis with automatic background fitting (Simmonds et al., 2018) was performed on the eROSITA spectra extracted in the 0.2-10 keV bands of all the eFEDS point-like sources with BXA (Buchner et al. 2014), which connects XSPEC (Arnaud 1996) with the UltraNest\footnote{\url{https://github.com/JohannesBuchner/UltraNest/}} nested sampling algorithm (Buchner 2016, 2019). 
Several models have been adopted (single and double power-law model, thermal model, and, for the sources with the lowest statistics, models with fixed photon index). All the details of the spectral fitting are reported in Liu et al. (submitted). 
The benefit of nested sampling includes that the procedure automatically characterizes all posterior modes, measures the uncertainties and multi-variate degeneracies, identifies its convergence automatically, and enables Bayesian model comparison among different models.
A main benefit for large surveys is that nested sampling can be run unsupervised on each source, regardless of spectral quality, and produces homogeneous output products.

Figure 2 shows the {\it eROSITA} total and background spectra, with the best fit model of an absorbed power law at the redshift of the source. 
Adopting a uniform prior for the slope and a logarithmic prior for N$_H$, the best fit photon index and column density are  $\Gamma$=2.27$^{+0.39}_{-0.44}$ and N$_H$=2.75$\pm{0.65}\times10^{22}$ cm$^{-2}$, respectively. 

We also fit the X-ray spectrum with an absorbed thermal model at the redshift of the source ({\tt tbabs*apec}), which returned a best fit temperature of kT$\sim3.5$ keV and a column density of $\sim8\times10^{21}$ cm$^{-2}$, with a statistical significance similar to the absorbed power-law fit.  However, these physical properties are hardly justified when compared to other properties of the system (see Section 2.2). We therefore consider the absorbed power-law model the one which best reproduces our data.  

The intrinsic 2-10 keV luminosity from the best fit model is L$_X=6.5^{+1.15}_{-1.0}\times10^{44}$ erg s$^{-1}$. From an X--ray point of view the source is therefore classified as an obscured QSO, confirming the effectiveness of the selection criterion in revealing these rare systems.

\begin{figure*}
\centering
  \includegraphics[width=18cm]{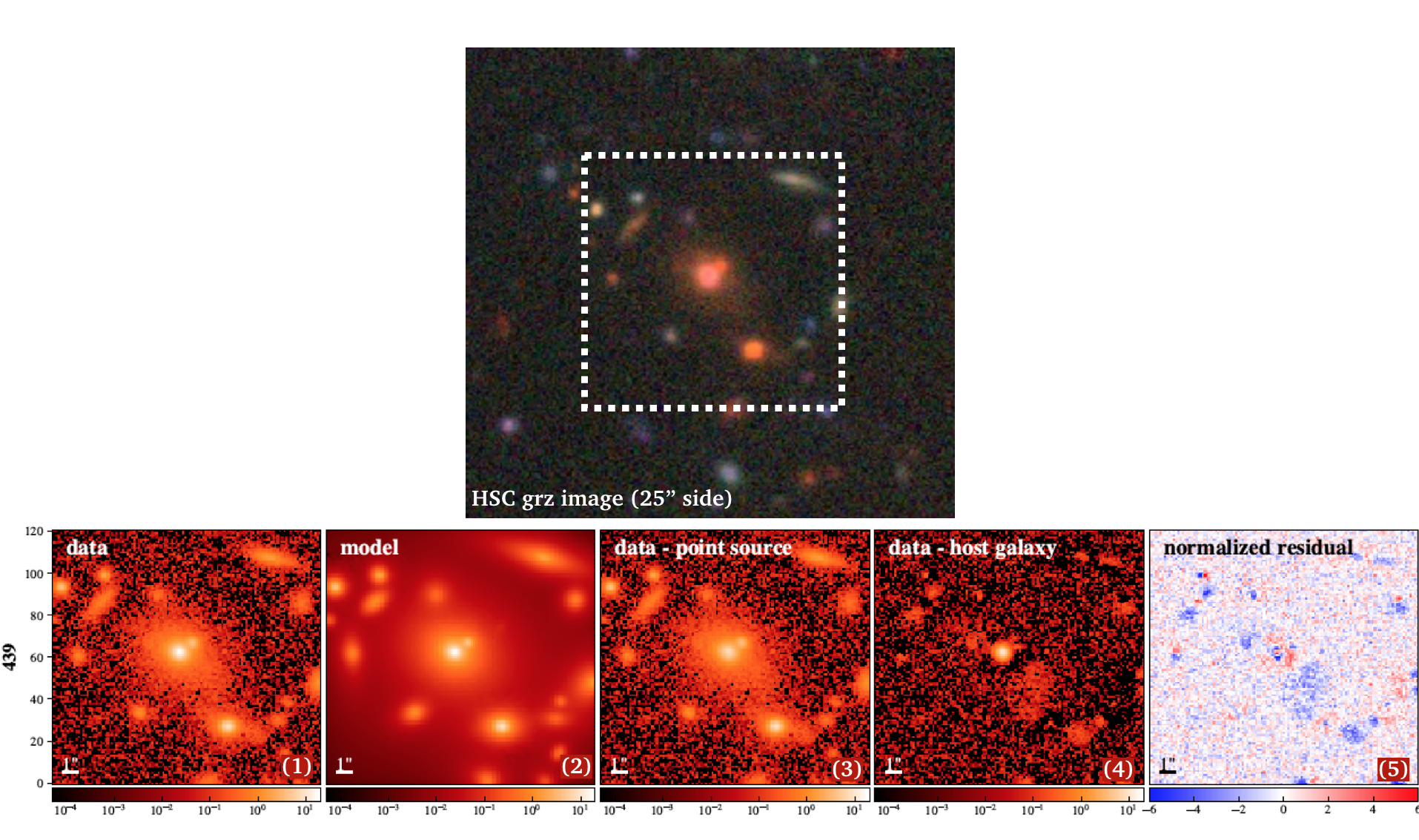} 
\caption{
{\it Top panel}: HSC color image (band: grz) of the windy QSO XID439. The cutout size is $\sim$25\arcsec on side. The dashed rectangle highlights the area where the host galaxy-AGN decomposition has been performed. {\it Bottom panel}: Host galaxy-AGN decomposition on Subaru i-band data. The panels from left to right are a zoom in the central area of the top panel (dashed rectangle, $\sim$10\arcsec in size), as follows: (1) observed HSC i-band image; (2) best-fit point source + galaxy model; (3) data minus the point source model (i.e., the pure-galaxy image); (4) data minus the galaxy model (i.e., the pure-AGN image); (5) fitting residual divided by the variance map. In all panels, North is up and East is left.} 
\label{HSC}
\end{figure*}

\subsection{Host galaxy properties}
The HSC {\it grz} color image is shown in the top panel of Fig. 3. XID439 is the red, bright source located at the center (r=20.74). An apparent over-density of sources is visible around the target, including a close-by, fainter (r=22.53) object at 0.5-1\arcsec in the NW direction, possibly signalling the presence of an ongoing merger. 

The observed X-ray fluxes for XID439 are well above the confusion limit expected for eROSITA in the soft and hard band ($\sim2\times10^{-15}$ erg cm$^{-2}$ s$^{-1}$ and $\sim1\times10^{-14}$ erg cm$^{-2}$ s$^{-1}$, respectively; see Kolodzig et al. 2013). We can therefore argue against the fact that the observed X-ray emission is the superposition of two sources. 
Moreover, the point-like classification in the {\it eROSITA} catalog and the hard X-ray spectrum reasonably exclude that the observed X-ray emission could be attributed to hot gas emission in a galaxy group. This is also confirmed by the fact that the best fit thermal model returns a best fit temperature of 3.5 keV (see Sect. 2.1), significantly larger than the temperature expected for galaxy groups (kT$<$ 3 keV; e.g. Eckmiller et al. 2011). 
Also, the best fit thermal model requires a large column density, which is hard to physically motivate in extended source emission. 

From the broadband optical and NIR photometry available from HSC and WISE 
(Wright et al. 2010), we were able to decompose the nuclear and host galaxy emission (Li et al. submitted). 
The point source is detected in all HSC bands, however the overall emission is dominated by the host galaxy (with host-to-total flux ratio of $\sim$80\%), with a half light radius of r$_e$=1.4 arcsec ($\sim 9.5$ kpc). 
Figure~\ref{HSC}, bottom panel, shows the HSC data in the I-band, and the results of the host-galaxy/AGN decomposition (see caption for details). 
After accounting for the host-galaxy and point-like emission, residual emission on scales of the order of 1\arcsec in the central region of the galaxy are still present (rightmost panel in Fig. 3).  

We run \texttt{X-CIGALE} (Boquien et al. 2019, Yang et al. 2020) on the entire SED, including also the MIR and FIR datapoints and upper limits, and the X-ray {\it eROSITA} fluxes. 
The parameters set-up for our reference model is illustrated in Appendix A.1, and it is the same used for the fit of the WISE selected sources in the eFEDS field (Toba et al. submitted). 

The stellar mass is constrained to be M$_*=$4.4$\pm1.4\times10^{11}$ M$_\odot$, and for the SFR we retrieve SFR=62$\pm$20 M$_\odot$ yr$^{-1}$, although weakly constrained due to the upper limits in the Herschel bands (see Fig. A.1and discussion in the Appendix). The galaxy is not classified as a starburst galaxy but rather it could be classified as a main sequence or sub-main sequence galaxy at its redshift.
 
We also estimated the 6 $\mu$m luminosity contributed from AGN to be $L_6^{AGN} = (2.78 \pm 0.39) \times 10^{45}$ erg s$^{-1}$ in the same manner as \cite{Toba_19}. 
If we assume an empirical relation between $L_6^{AGN}$ and $L_{\rm 2-10 keV}$ \citep{Chen}, the expected $L_{\rm 2-10 keV}$ is $\sim5 \times 10^{44}$ erg s$^{-1}$, consistent with the value obtained from our X-ray spectral fitting (Sect. 2.1). This is a further, independent confirmation that the X-ray flux may be associated to the point-like AGN.

   \begin{figure*}
   \centering
  \includegraphics[width=18cm]{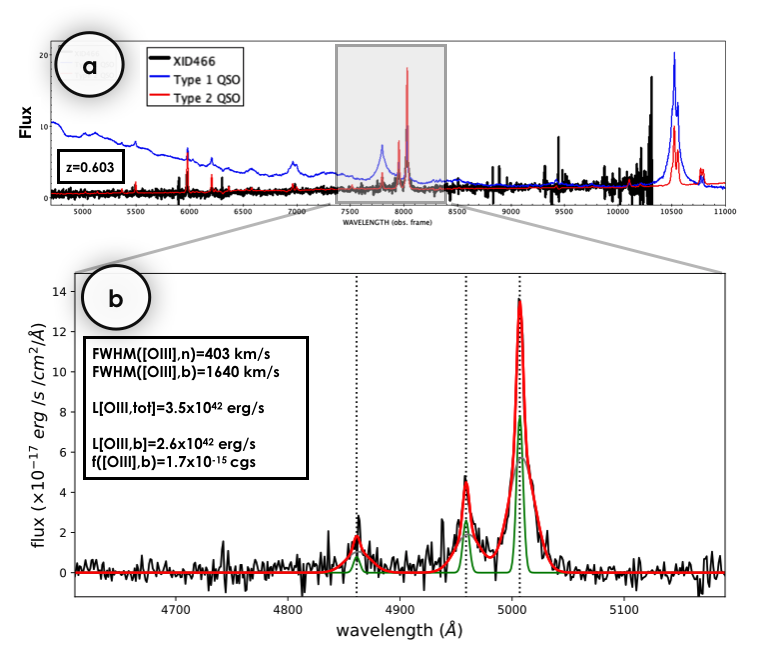}
      \caption{SDSS spectrum of XID439 (panel a) from $\sim5000$ to $\sim$10000 \AA\, with superimposed the templates of Type 1 and Type 2 AGN, as labeled, highlighting the significant  reddening of XID439 and its likely Type 2 nature.    A   close-in zoom of the H$\beta$+[OIII] lines complex is shown (panel b) with the 2 gaussians best fit decomposition.      }
         \label{FigFit}
   \end{figure*}

\section{Results}

\subsection{An ionised outflow}

The SDSS spectrum\footnote{http://skyserver.sdss.org/dr16/en/tools/explore/summary.aspx?plate=\\ 3819\&mjd=55540\&fiber=65} is shown in Fig. 4a, compared with templates of Type 1 and Type 2 AGN (Vanden Berk et al. 2001, Yuan et al. 2016). The Type 2 template seems to fit the overall spectrum (continuum and lines) reasonably well. We note that a similar fit can be obtained with a considerable extinction, E(B-V)$\sim$0.6, applied to the Type 1 template (see Urrutia et al. 2009).  
A NIR spectrum sampling the H$\alpha$ would be key to disentangle the two scenarios. 

The spectrum unambiguously shows the presence of broad [O III]4959,5007 lines: a non parametric analysis returns a total line profile width of w80=1560 km s$^{-1}$, significantly larger than the median value of $<w80>_{med}\sim700$ km s${^-1}$ obtained for the Type 2 quasar sample from SDSS presented in Yuan et al. (2016).

In orther to further constrain the line properties, we performed a multi-component Gaussian fit,
 which required the presence of two gaussian lines with FWHM$\sim$400 km s$^{-1}$ and FWHM$\sim$1650 km s$^{-1}$, respectively. Figure 4b shows a close-up of the H$\beta$ and [O III]4959,5007 region with the two components shown separately. The two components fit was preferred over the one component (that would return a FWHM=1080 km s$^{-1}$) on the basis of the BIC criterion (Schwarz 1978). 
The broad component is slightly redshifted ($\sim 100$ km s$^{-1}$). Although less frequent than the blueshifted cases, redshifted outflows are still common in Type 2 AGN samples showing disturbed [OIII] kinematics (e.g.  Bae \& Woo 2014, Yuan et al. 2016, Perna et al. 2017).

A combination of two gaussian lines with the same widths and shifts that fit the [OIII]4959,5007 doublet can also reproduce the weak H$\beta$4861 emission. In this scenario, the broad component seen in H$\beta$ can therefore be associated with an ionised wind rather than to  Broad Line Region (BLR) motion, pointing towards a Type 1.9-2 nature for this source.

The observed luminosity associated with the outflowing broad component L$_{[OIII],broad}$ is $\sim$ 2.6$\times10^{42}$ erg/s, one to two orders of magnitude larger than that observed  with IFU in local Seyferts (e.g. MAGNUM and CARS samples; Mingozzi et al. 2019, Powell et al. 2018) 
and comparable to that observed in a sample of  z$\sim0.5$ Type 1 QSOs (e.g.  Husemann et al. 2016). 

Following Fiore et al. (2017), we computed the ionised gas mass outflow rate ($\rm \dot{M}_{ion}$) assuming  v$_{out}=\Delta V+2\sigma_{broad}\sim1400$ km s$^{-1}$ as outflow velocity, the observed luminosity of the broad [OIII] component, and adopting for the electron density of the medium a value of n$_e$=500 cm$^{-3}$ (see also Kakkad et al. 2020). For the spatial extension of the outflow we assume $R_{out}\sim$10 kpc, corresponding to the half light radius measured from HSC. We obtain $\rm \dot{M}_{ion}\sim1.4^{+6.2}_{-1.2}$ M$_\odot$/yr, where the $\dot{M_{ion}}$ uncertainties are obtained adopting a Monte-Carlo approach, following Marasco et al. (2020; see Appendix B).  
This mass outflow rate should be considered as a lower limit given that no correction for extinction has been adopted, and we assumed as spatial extension the entire galaxy scale. Given that the mass outflow rate is directly proportional to the observed luminosity and inversely proportional to the extention (see eq. B.2 and B.3 in Fiore et al. 2017), the mass outflow rate can be up to $\sim 2$ orders of magnitudes larger in case it is confined only within the central 1 kpc and assuming a correction of 10 for the extinction.

 \subsection{AGN properties and Eddington ratio}

 Kim et al. (2015) reported an estimate of the bolometric luminosity (L$_{bol}\sim1.5\times10^{46}$ erg s$^{-1}$), BH mass (M$_{BH}$ = 2.8$\times10^{8}$ M$_{\odot}$) and Eddington ratio (L$_{bol}$/L$_{Edd}$=0.486) for  XID439, as derived from the Pa$\beta$ line in the framework of NIR spectroscopic follow-up of the sample of Red QSOs presented in Urrutia et al. (2009). 
  We derived a more accurate estimate of the bolometric luminosity from the SED fitting with \texttt{X-CIGALE}, obtaining L$_{bol,AGN}=7.8^{+1.2}_{-4.8}\times10^{45}$ erg s$^{-1}$ (see also Appendix A), a factor of $\sim2$ lower than that reported in Kim et al. (2015). This  translates into a corresponding lower estimate for the Eddington ratio, L$_{bol}$/L$_{Edd}$=0.25.
 
This value is more than one order of magnitude higher than the value L$_{bol}$/L$_{Edd}\sim$0.01, inferred for the X--ray selected AGN population at z$\sim$0.5-1.0 as reported in Georgakakis et al. (2017) and calculated from the ratio between the X--ray luminosity and  the host galaxy stellar mass (known as ``specific accretion rate", see also Brusa et al. 2009), after an assumption on the bolometric correction between the X--ray and the bolometric luminosity (k$_{bol,X}$=L$_{bol}/L_{2-10 keV}$=25), and on the M$_{BH}$/M$_*$ ratio (M$_{BH}\sim0.002\times M_*$ e.g Marconi \& Hunt 2003)%
\footnote{This method, with the same assumptions, is widely used in the literature to determine the Eddington ratio for large samples of X-ray selected AGN, when SED fitting and/or optical/NIR spectroscopy sampling broad emission lines are not available (e.g. Brusa et al. 2009). With this method, we would obtain for XID439: L$_{bol}\sim1.6\times10^{46}$ erg s$^{-1}$, M$_{BH}\sim 8.8\times10^{8}$ M$_{\odot}$ and L$_{bol}$/L$_{Edd}\sim$0.1, still significantly larger than the average value. }. 
This is a further confirmation that XID 439 is experiencing a peculiar phase, and it has been caught close to the most active, accreting phase.  We note however that the high value observed is mostly due to the high X--ray luminosity of XID439. Indeed, further downselecting all AGN with L$_X>10^{44.5}$ erg s$^{-1}$ (larger than the eFEDS limit) in the same redshift range we obtain a consistent value (L$_{bol}$/L$_{Edd}=0.25^{+0.25}_{-0.12}$, $\sim$ 1$\sigma$ errors).  On the other hand, the population of sources with L$_{bol}$/L$_{Edd}>0.25$ has, on average, a significantly lower luminosity (logL$_{X}=43.7^{+0.64}_{-0.45}$) 
and stellar mass (logM$_*$= 10.13$^{+0.6}_{-0.55}$)  than those observed in XID439, pointing towards the fact that that the proposed selection is efficient in isolating massive, highly accreting SMBHs.

All properties discussed in the text and derived from the data presented in this work are listed in Table~\ref{table:1}.

\section{Discussion}

XID439 satisfies several selection criteria used in the past to isolate obscured QSOs.
It has been reported for the first time in the sample of 229 optically Type 2 QSOs at z$<0.83$ from the SDSS survey on the basis of a high [OIII]5007/H$_\beta$ flux ratio (Zamaska et al. 2003; see also Reyes et al. 2008).  The sample has been enlarged subsequently in the work by Yuan et al. (2016), which contains 2920 Type 2 QSOs out to z$\sim$1\footnote{XID439 is also in the extremely red quasar sample of 645 objects selected from the cross-correlation of SDSS, BOSS and WISE catalogs, on the basis of a red r-W4$>$14 (Ross et al. 2015): of the 2 sources falling in the eFEDS area, XID439 is the only one detected by {\it eROSITA.}}. Fourty-three out of 2920 objects fall within the eFEDS footprint. Of these, only 2 are significantly detected by {\it eROSITA} in the 2.3-5 keV band.
The detected sources increase to 6 when the full (0.2-5 keV selected) eROSITA catalog is considered.
XID439 is the most X-ray  luminous one, and the only with a line width (W80, see Sect. 3) $>800$ km s$^{-1}$.
We did not reveal significant X-ray emission in the eROSITA stacking of the remaining undetected 37 sources, at an average redshift of z$\sim$0.5 (see Appendix C for details). This corresponds to an average 2-10 keV 
luminosity lower than $\sim2\times10^{43}$ erg s$^{-1}$ (unless all these objects are  heavily obscured).  The average hard X-ray luminosity is consistent with that  expected for Type 2 Seyfert galaxies of comparable [OIII] luminosities (L$_{[OIII]}\sim5\times10^{42}$ erg s$^{-1}$), according to the L$_X$-L$_[OIII]$ relation of Heckman et al. (2005). 

\begin{table}
\caption{Target properties.}           
\label{table:1}      
\begin{tabular}{l r}        
\hline\hline                
Name & eFEDSU J091157.5+014327$^1$\\ 
RA (J2000) &   09:11:57.557 \\
DEC (J2000) & +01:43:27.54   \\
zspec & 0.603  \\ 
 M$_{\rm BH}$/M$_\odot$ & 2.8$\times10^8$\\
\hline
F$_{\rm 2.3-5 keV}$/erg cm$^{-2}$ s$^{-1}$ & 2.46$\times10^{-13}$ (catalog) \\
F$_{\rm 2-10 keV}$/erg cm$^{-2}$ s$^{-1}$  & 3.68 $\times10^{-13}$  (spectral fit)\\
L$_{\rm 2-10 keV}$/ erg s$^{-1}$\  &   $6.5^{+1.15}_{-1.0}\times10^{44}$\\
N$_{\rm H}$/cm$^{-2}$   & 2.75$\pm{0.65}\times10^{22}$ \\
$\Gamma$ & 2.27$^{+0.39}_{-0.44}$ \\
 \hline
r (LS8) & 20.74 \\  
r (HSC, host) & 20.85 \\ 
W1 & 15.94 \\
\hline
r$_e$ & 1.4\arcsec (9.5 kpc)\\
M$_{\star}$/M$_\odot$  & 4.4$^{+7.6}_{-1.4}\times10^{11}$ \\
SFR (SED)/M$_\odot$/yr & 2$^{+68}_{-1.8}$ \\
\hline
FWHM[OIII]$_{broad}$/km s$^{-1}$  & $1640_{-30}^{+42}$\\ 
FWHM[OIII]$_{narrow}$/km s$^{-1}$ & $403_{-12}^{+17}$ \\ 
w80[OIII]/km s$^{-1}$ & 1560$\pm56$ \\ 
v$_{out}$/km s$^{-1}$ & 1400 \\ 
L$_{\rm [OIII]}$/erg s$^{-1}$  & 3.5$\pm0.04\times10^{42}$\\ 
L$_{\rm [OIII],broad}$/erg s$^{-1}$  & 2.6$\pm0.05\times10^{42}$\\  
\hline
$\dot{M_{ion}}$/M$_\odot$ yr$^{-1}$  & $1.4_{-1.2}^{+6.2}$ \\   
L$_{\rm bol, AGN, SED}$/erg s$^{-1}$  & 7.8$^{+1.2}_{-4.8}\times10^{45}$ \\
k$_{\rm bol, HX}$  & 12$^{+4}_{-8}$ \\
L/L$_{\rm Edd}$ & 0.25  \\
\hline                                 
S$_{1.4 GHz}$/mJy (FIRST) & 4.56$\pm0.135$ \\
L$_{1.4 GHz}$/W Hz$^{-1}$  & 24.83 \\
\hline
\end{tabular}
\\
Notes: $^1$ This source is referred in the text as XID439 and is also known as: F2MS 0911+0143; SDSS J091157.54+014327.6; $^2$ BH mass value from Kim et al. (2015). The uncertainties associated to M$_\star$, SFR, L$_{\rm bol}$ and k$_{\rm bol, HX}$ reflect the degeneracies in the SED fitting (see Appendix A.1).
\end{table}

Our target is also part of the FIRST-2MASS (F2MS) sample of 122 radio-selected QSOs with red colors (R-K$>4.5$ and J-K$>1.3$) presented in Urrutia et al. (2009; F2MS 0911+0143).  Only 4 objects from this sample fall within the eFEDS footprint, 
XID439 is the only detected by {\it eROSITA} in the hard band. 
It has a peak flux at 1.4 GHz of 4.5 mJy as measured from FIRST (Becker et al. 1995), where it remains unresolved. It is not a strongly jetted source and its radio-loudness parameter, {\it R}, is  $<10$, so it is classified as a radio-quiet source, with a spectral index of -0.74 between 1.4 and 0.325 GHz (Mauch et al. 2013).  The radio power of L$_{\rm 1.4 GHz}\sim10^{25}$ W Hz$^{-1}$ and the steep spectral index would classify XID439 as a Compact Steep Spectrum radio source, expected to be intrinsically young sources (e.g. Fanti et al. 1995), further pointing toward the hypothesis that XID439 may be caught in an early stage of its evolution. 

Finally, XID439 is part of the WERGs (Wide and Deep Exploration of Radio Galaxies with Subaru HSC, Yamashita et al. 2018) sample presented in Toba et al. (2019), which contains 1056 radio galaxies at z$<1.7$ selected from FIRST and matched with HSC down to g$\sim$26, with accurate SED fitting\footnote{In that catalog an incorrect redshift was used for the SED fitting of XID439 and the derived parameters are superseded by this work.}. 425 sources lie in the eFEDS footprint: this number is considerably larger than those listed above, given that no further optical-to-IR color pre-selection has been applied to the sample. Ichikawa et al. (in prep.) will report the complete analysis of the X--ray properties of WERGS in eFEDS. 

Both the red QSOs selections and the WERGS survey have revealed optically-faint radio galaxies and have been proved to be particularly effective in selecting the high Eddington ratio AGN population based on observed source properties (e.g. color and/or flux cuts). 
Our selection based on a combination of X-ray to optical and optical to NIR colors is similarly very effective in detecting such highly accreting, obscured sources. Although initially proposed in a relatively small sky area such as COSMOS observed with XMM-Newton (Brusa et al. 2010; see also La Massa et al. 2016 for an extension to Stripe82), we have probed that it can be applied also at much brighter fluxes and on samples extracted from the {\it eROSITA} hard X-ray surveys.  
Indeed, the only source (out of 3 candidate outflowing QSOs) with an available optical spectrum, XID439, is an X-ray Type 2 QSO with  a robustly constrained L/L$_{Edd}$=0.25 from accurate SED fitting and available BH mass (Section 3.2). 

Despite XID439 was already present in several samples in the past from optical or radio selections (all based on all-sky survey area or, at minimum, area covered by the SDSS footprint), it has never been followed-up in the X-rays before. {\it eROSITA} provides therefore the first X-ray observation ever of this object, confirming its prominent X-ray luminosity and obscuration (N$_{\rm H}>10^{22}$ cm$^{-2}$). We note that also the other two sources in the outflowing locus are X-ray obscured with N$_H>10^{22}$ cm$^{-2}$ and intrinsic L$_{2-10 keV}\sim10^{44.5,45.5}$ erg s$^{-1}$ when the photometric redshfit are considered, z$_{phot}$=0.8,2.2 (Liu et al. submitted).
Moderate X--ray obscuration in the Compton thin regime has been reported also from previous X-ray follow-up of red QSOs at similar redshift and luminosities (Wilkes et al. 2002, Urrutia et al. 2005, Brusa et al. 2005). At higher redshift and higher luminosity, instead, the population of extremely red QSO shows on average higher column densities (N$_{\rm H}\sim10^{23}$ cm$^{-2}$), suggesting that these objects are partially hidden by their own equatorial outflows with a large opening angle (Goulding et al. 2018). 

Sources accreting close to the Eddington limit with significant X-ray obscuration are expected to be caught in the feedback phase of merger based galaxy-AGN coevolutionary models (Fabian et al. 2008, Kakkad et al. 2016, Lansbury et al. 2020). This has been further confirmed by an unambiguous signature of an ionised wind in the [OIII] emission line profile (see Sect. 3.1), which remained unnoticed in all previous works.  The derived outflow properties are in line with results from samples of z$\sim2$ unobscured AGN (e.g. Kakkad et al. 2020). 
The relatively closer distance with respect to its z$>1$ analogous makes eFEDSU J091157.5+014327 a perfect target to further advance in the spatially resolved studies of ionised winds in {\it obscured QSOs} and their interaction with the ambient, via dedicated MUSE and/or JWST observations. 

   \begin{figure}
   \centering
  \includegraphics[width=9cm]{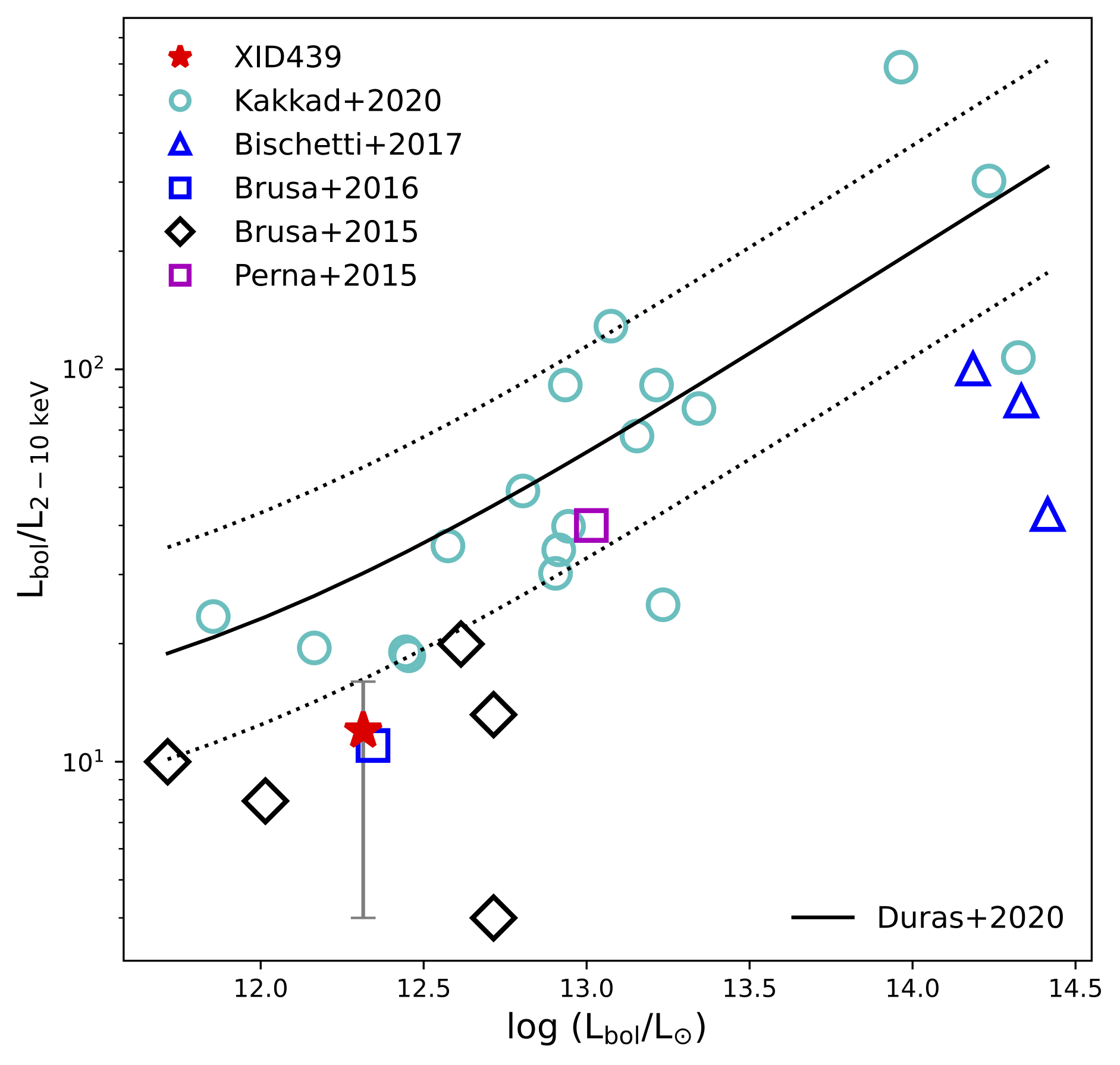} 
      \caption{Hard X-ray Bolometric correction as a function of the bolometric luminosity. XID439 is indicated by the red star. Other X-ray QSOs at \mbox{z$\sim1-3.5$} for which ionised gas outflows on kpc-scale have been detected and for which a solid estimate of the bolometric luminosity from SED fitting is available are also plotted, as labeled. In most cases both X-ray spectral analysis and SED fitting has been performed on data of similar quality (in terms of counting statistics, depth of the images, photometric errors) of the analysis of XID439 and we expect that the associated errorbar would be similar.
      The solid and dotted lines are the analytic expressions for the hard X-ray bolometric correction and its scatter derived in Duras et al. (2020).}
         \label{figkbol}
   \end{figure}

The direct comparison between the AGN bolometric luminosity obtained from the SED fitting and the hard X-ray luminosity measured by {\it eROSITA} implies an hard X--ray bolometric correction k$_{bol,HX}\sim$12 for our source. This value is a factor of $\sim$2 lower than the one expected at similar luminosities from the correlation between the bolometric luminosity and the hard X-ray bolometric correction as reported in Duras et al. (2020; see also Lusso et al. 2012).  
Figure~\ref{figkbol} shows the position of XID439 in the k$_{bol, HX}$ vs. L$_{bol, AGN}$ plane, compared  with the Duras et al. (2020) relation. The error bar comprises our uncertainties in the estimate of the bolometric luminosities from the SED fitting (see Appendix) and the uncertainties on the X-ray luminosity. We also report the values derived for a small sample of other X-ray obscured, radio-quiet QSOs at z$\sim1-4$ in which kpc-scale ionised outflows have been revealed and for which the bolometric luminosity has been derived from an accurate SED fitting (see references in the figure). Interestingly, for this small sample, k$_{bol, HX}$ appears  lower than the average value at a fixed L$_{bol, AGN}$: in particular, we note that 12/29 sources have k$_{bol, HX}$ lower than the $-$1$\sigma$ scatter of the correlation, to be compared to only 2 with k$_{bol, HX}$ higher than the +1$\sigma$ scatter. 
 A higher contribution of L$_{X}$ to L$_{bol}$ may in principle be ascribed to the presence of an inner ADAF and an outer truncated accretion disc as observed for local low-luminosity AGN (see e.g. Qiao et al. 2013). However, this may not be the case for our sources with kpc-scale ionised outflows, all with bolometric luminosities larger than 10$^{45}$ erg s$^{-1}$.
Giustini \& Proga (2019) proposed that the diversity of the accretion and ejection flows in AGN can be explained, in addition to the variation of the Eddington ratio and the black hole mass, by the inclusion of accretion disc winds. 
The ionised winds observed at kilo-parsec scales are expected to be a subsequent phase of the evolution of accretion disc winds. 
Although the statistics is still limited, the fact that the sources are caught in a peculiar phase of their evolution can be the reason of the observed atypical balance of disk and corona emissions with respect to the overall population of Type 1 and Type 2 QSO (see also Brusa et al. 2016, Perna 2016). Larger samples of quasars with ionised outflows and well constrained bolometric correction will result critical in testing this scenario.

Finally, the combined X-ray, radio and red color selection represents the best approach to pick-up luminous AGN with all the expected properties of objects caught in the short-lived, feedback phase. 
The observed radio emission can signal the presence of relativistic, unresolved jets in young radio sources which may be responsible of most of the ejection of material (see e.g. Jarvis et al. 2019). 
The availability for the first time of a hard X--ray selected AGN sample over the entire sky, with associated  spectroscopic coverage, will be crucial to investigate the existence of ionised, outflowing gas  also in X-ray sources without prominent radio emission. 
The results presented in this work therefore constitute a  pathfinder for {\it eROSITA} discoveries in the years to come over the full sky. By scaling the eFEDS area to the total extragalactic sky (a factor of $\sim$250 in area), we expect a total number of $\sim750$ outflowing QSOs candidate in the {\it eROSITA} hard band catalog. Half of them will lie in the German part of the sky, where spectroscopic follow-up by SDSS-V (Kollmeier et al. 2017) and 4MOST (de Jong et al. 2019) will be available and will provide immediate spectroscopic identification for all sources brighter than r=22.5. 
Similarly, sources detected in the {\it eROSITA} all-sky hard X-ray survey and matched with adeguate spectrosopic information will enable the discovery of larger samples of (obscured) QSOs with simultaneous low k$_{bol}$ and clear outflow signatures, needed to constrain feedback models.

\begin{acknowledgements}
This work is based on data from eROSITA, the soft X--ray instrument aboard SRG, a joint Russian-German science mission supported by the Russian Space Agency (Roskosmos), in the interests of the Russian Academy of Sciences represented by its Space Research Institute (IKI), and the Deutsches Zentrum f\"ur Luft- und Raumfahrt (DLR). The SRG spacecraft was built by Lavochkin Association (NPOL) and its subcontractors, and is operated by NPOL with support from the Max-Planck Institute for Extraterrestrial Physics (MPE).
The development and construction of the eROSITA X-ray instrument was led by MPE, with contributions from the Dr. Karl Remeis Observatory Bamberg \& ECAP (FAU Erlangen-Nuernberg), the University of Hamburg Observatory, the Leibniz Institute for Astrophysics Potsdam (AIP), and the Institute for Astronomy and Astrophysics of the University of T\"ubingen, with the support of DLR and the Max Planck Society. The Argelander Institute for Astronomy of the University of Bonn and the Ludwig Maximilians Universit\"at Munich also participated in the science preparation for eROSITA.
The eROSITA data shown here were processed using the eSASS/NRTA software system developed by the German eROSITA consortium.
The Hyper Suprime-Cam (HSC) collaboration includes the astronomical communities of Japan and Taiwan, and Princeton University.  The HSC instrumentation and software were developed by the National Astronomical Observatory of Japan(NAOJ), the Kavli Institute for the Physics and Mathematics of the Universe (Kavli IPMU), the University of Tokyo, the High Energy Accelerator Research Organization (KEK), the Academia Sinica Institute for Astronomy and Astrophysics in Taiwan (ASIAA), and Princeton University.  Funding was contributed by the FIRST program from Japanese Cabinet Office, the Ministry of Education, Culture, Sports, Science and Technology (MEXT), the Japan Society for the Promotion of Science (JSPS), Japan Science and Technology Agency (JST),the Toray Science Foundation, NAOJ, Kavli IPMU, KEK,ASIAA, and Princeton University.
Funding for the Sloan Digital Sky Survey (SDSS) has been provided by the Alfred P. Sloan Foundation, the Participating Institutions, the National Aeronautics and Space Administration, the National Science Foundation, the US Department of Energy, the Japanese Monbukagakusho, and the Max Planck Society. The SDSS Web site is http://www.sdss.org/. The SDSS is managed by the Astrophysical Research Consortium (ARC) for the Participating Institutions. The Participating Institutions are The University of Chicago, Fermilab, the Institute for Advanced Study, the Japan Participation Group, The Johns Hopkins University, Los Alamos National Laboratory, the Max-Planck-Institute for Astronomy (MPIA), the Max-Planck-Institute for Astrophysics (MPA), New Mexico State University, University of Pittsburgh, Princeton University, the United States Naval Observatory, and the University of Washington.
MB acknowledges support from PRINMIUR2017PH3WAT  (‘Black  hole  winds  and  the  baryon  life  cycle of galaxies’). MP is supported by the Programa Atracci\'on de Talento de la Comunidad de Madrid via grant 2018-T2/TIC-11715, and the Spanish Ministerio de Econom\'ia y Competitividad through the grant ESP2017-83197-P, and PID2019-106280GB-I00. BM is supported by the European  Innovative Training  Network  (ITN)  "BiD4BEST" funded  by  the  Marie  Sklodowska-Curie Actions in Horizon 2020 (GA 860744). JW acknowledges support by the Deutsche Forschungsgemeinschaft (DFG, German Research Foundation) under Germany’s Excellence Strategy - EXC-2094 -390783311.

\end{acknowledgements}

\begin{appendix} 
\section{Spectral Energy Distribution}

The parameter ranges used in the SED fitting are summarized in Table A.1, in which we modeled the SED as follows.
We assumed a delayed star formation history (SFH).
For dust attenuation, we used a starburst attenuation curve provided by \cite{Calzetti}.
We model the photometry with a single stellar population (SSP) template from the \cite{Bruzual} library, assuming the initial mass function (IMF) of \cite{Chabrier}, and the standard nebular emission model included in {\tt X-CIGALE} \citep[see ][]{Inoue}.
AGN emission was modeled using a clumpy two-phase torus model \citep[{\tt SKIRTOR}:][]{Stalevski}.
We used the dust emission model provided by \cite{Draine}.
X-ray emission was modeled with fixed power-law photon indices of AGN, low-mass X-ray binaries (LMXBs), and high-mass X-ray binaries (HMXBs).
A full explanation of the photometry used for the SED fitting and SED modeling for WISE-detected X-ray sources including XID439 is given by \cite{Toba_21a,Toba_21b}.

In figure A.2 we present the SED fitting to a total of 19 photometric points and upper limits, from the X--ray to far infrared available for XID439. From this best fit solution, for the present work, we adopt the values of L$_{\rm bol, AGN}$, M$_{\star}$, and SFR.\\

We further checked for internal degeneracies in the SED fitting components, by relaxing some of the assumed input parameters (e.g. the inclination disk angle $\Theta$, the maximum allowed range for the X-ray photon index). 
We acknowledge that we can reproduce the observed SED with several combination of the best fit parameters, with comparable statistical significance. However, in all the best fit solutions we recover values for the stellar mass (3-12$\times10^{11}$ M$_{\odot}$) and AGN bolometric luminosities (3-9$\times10^{45}$ erg s$^{-1}$) bracketing our best fit values in the reference set-up. 
For the SFR we retrieved a full range of values of 0.2-70 M$\_{\odot}$ yr$^{-1}$, with a median value of 2 M$\_{\odot}$ yr$^{-1}$. As already noted in Section 2.2, a solid estimate of the SFR is basically limited by the non detection in the Herschel bands. In any case, there seems to be no significant evidence for a recent starburst in any setup configuration. The fact that a recent starburst can be excluded is a further indication that we are catching the source in the "feedback" phase (expected to follow the starburst one).

\begin{figure*}
\includegraphics[width=18cm]{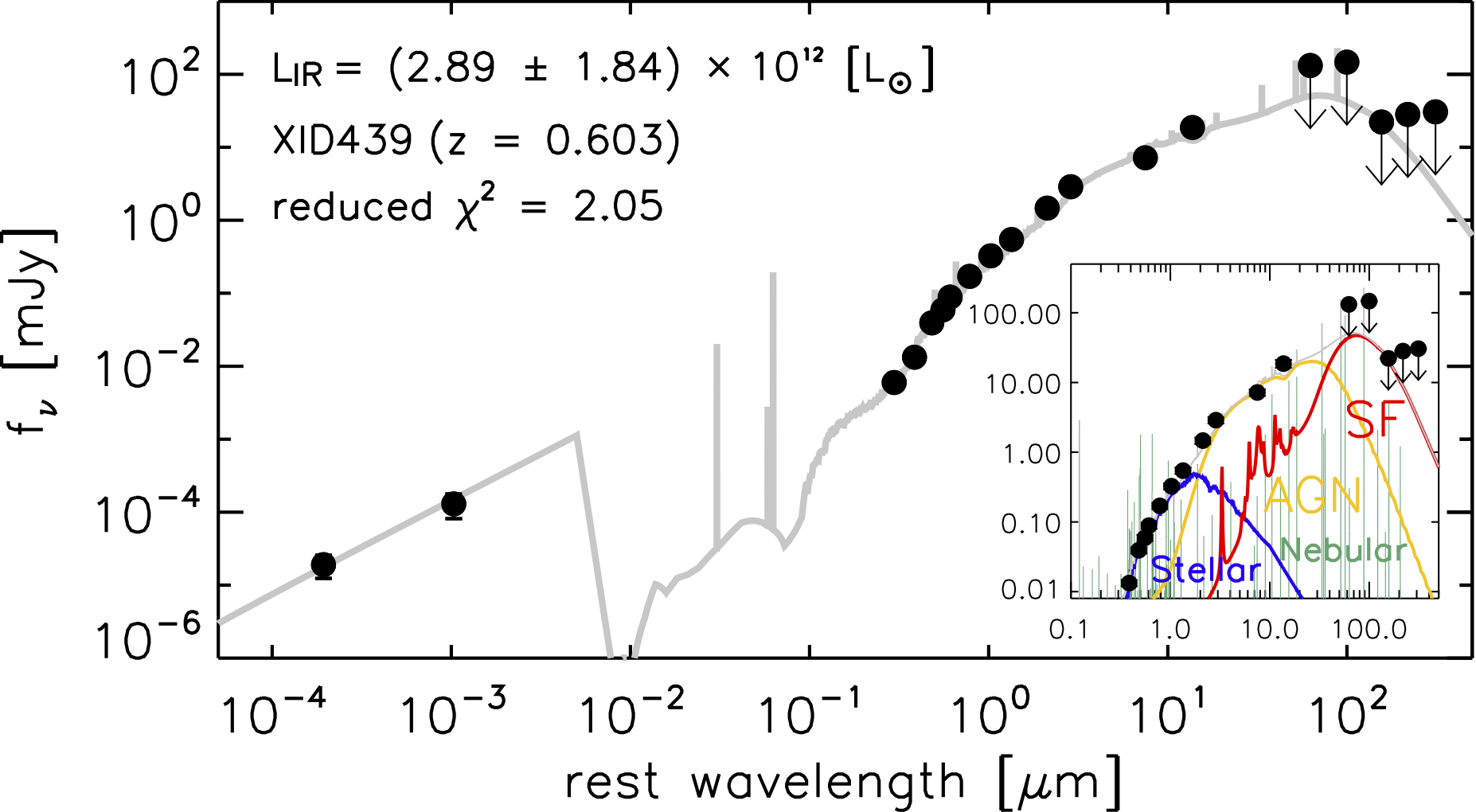}
\caption{
The best-fit SED from X-ray to FIR of eFEDSU J091157.5+014327 obtained using \texttt{X-CIGALE}. The black points are all the photometric data and the gray solid line represents the best-fit SED. The inset figure shows the SED at 0.1--500 $\mu$m where the contributions from the stellar, nebular, AGN, and SF components to the total SED are shown, as labeled.} 
\end{figure*}

\begin{table}
\caption{\bf Parameter ranges used in the SED fitting with \texttt{X-CIGALE}}
\label{Param}
\centering
\begin{tabular}{l c}
\hline \hline
Parameter & Value\\
\hline
\multicolumn{2}{c}{Delayed SFH}\\
\hline
$\tau_{\rm main}$ [Myr] & 1000, 4000, 8000, 12000 \\
age [Myr] & 500, 1000, 1500, 2000, 4000 \\
\hline
\multicolumn{2}{c}{SSP \citep{Bruzual}}\\
\hline
IMF				&	\cite{Chabrier} \\
Metallicity		&	0.02  \\
\hline
\multicolumn2c{Nebular emission \citep{Inoue}}\\
\hline
$\log\, U$	&	-2.0	\\
\hline
\multicolumn2c{Dust attenuation \citep{Calzetti}}\\
\hline
$E(B-V)_{\rm lines}$ &  0.05, 0.1, 0.5, 1.0, 1.5, 2.0 \\
\hline
\multicolumn{2}{c}{AGN emission \citep{Stalevski}}\\
\hline
$\tau_{\rm 9.7}$ 			&  	3, 7, 11 		\\
$p$							&	0.5,  1.5	\\
$q$							&	0.5, 1.5	\\
$\Delta$ [$\degr$]			&	10, 40, 80				\\
$R_{\rm max}/R_{\rm min}$ 	& 	30 		\\
$\theta$ [$\degr$]			&	50, 70, 90		\\
$f_{\rm AGN}$ 				& 	0.1, 0.2, 0.3, 0.4, 0.5, 0.6, 0.7, 0.8, 0.9 	\\
\hline
\multicolumn{2}{c}{Dust Emission \citep{Draine}}\\
\hline
 $q_{\rm PAH}$ &  2.50, 5.26, 6.63, 7.32 \\
 $U_{\rm min}$ & 10.00, 50.00 \\
 $\alpha$ & 1.0, 1.5, 2.0 \\
 $\gamma$ & 0.01, 0.1, 1.0 \\
\hline
\multicolumn{2}{c}{X-ray Emission \citep{2020MNRAS.491..740Y}}\\
\hline
AGN photon index	&	2.27	\\
$|\Delta\,\alpha_{\rm OX}|_{\rm max}$	&	0.2	\\
LMXB photon index	&	1.56	\\
HMXB photon index	&	2.0		\\
\hline
\end{tabular}
\\
Notes: For a full description of each parameter, see  \cite{2019A&A...622A.103B,2020MNRAS.491..740Y,Toba_21a,Toba_21b}.
\end{table}

\section{Uncertainties in the estimate of the Mass outflow rate} 
Uncertainties on $\dot{M_{ion}}$ are obtained taking into account all the possible values in the ranges [$3-30$] kpc and [$120-2000$] cm$^{-3}$  for $R_{out}$ and $n_e$ respectively, according to measured quantities at high-z (e.g. Kakkad et al. 2020 and Cicone et al. 2015, for the outflow extension; Brusa et al. 2015 and Forster-Schreiber et al. 2019 for the electron density). For the outflow velocity, we consider the range [$100-1650$] km/s, where the minimum value represent the velocity shift of the outflow component with respect to the systemic, while the maximum value is the line width of the outflow component in Fig. 4b.

\section{X--ray stacking of Type 2 QSOs} 
We stacked the 37 SDSS Type 2 Quasars sources  (z=0.4-0.65) undetected by eROSITA in the eFEDS X-ray footprint in the following way. 
First, in order to properly characterize the background, we mask previously detected sources  in the X-ray images, assuming a 30\arcsec radius for point-like sources and the measured spatial extent for the extended sources in the Brunner et al. (submitted) catalog. The procedure is done on 4 bands: full (0.2-10 keV), soft (0.2-0.6 keV), mid (0.6-2.3 keV) and hard (2.3-5.0 keV). We then stacked the signal in all four bands by taking the mean of inverse exposure weighted 2'x2' cutouts centered on each of the 37 Type-2 QSO position, basically creating count-rate stacked images. To not overly weigh sources at the edges of the field, we required a minimum exposure time of 180 s, which cut the number of stacked sources to 35. We tried to detect the source by PSF-matching the center to the background, failing to exceed a S/N of 1.5 in all four bands (the mid-band had a count rate of 2.24$\pm$1.68$\times10^{-3}$ cts s$^{-1}$, with a S/N of 1.33 above the background which we deem insufficient for a significant measurement). Lastly, we determine the 3 sigma limiting fluxes above the background to be detected using appropriate energy conversion factors calculated using an absorbed power-law model with column density log(N$_H$/cm$^{-2}$)=20 and photon index  $\Gamma$=1.7. We therefore can place upper limits of 1.58$\times10^{-15}$ erg cm$^{-2}$ s$^{-1}$ and 1.12$\times10^{-14}$ erg cm$^{-2}$ s$^{-1}$ in the 0.6-2.3 keV and 2.3-5 keV bands, respectively. 

\end{appendix}
\end{document}